\documentclass[manuscript]{acmart}
\renewcommand\footnotetextcopyrightpermission[1]{} 
\AtBeginDocument{%
  }

\usepackage{listings}
\usepackage{xcolor}
\lstdefinelanguage{none}{
  keywords={},
  keywordstyle=\color{black},
  sensitive=false,
  comment=[l]{//},
  morecomment=[l]{//}
}

\lstdefinestyle{promptblock}{
  language=none,
  basicstyle=\ttfamily\scriptsize,
  breaklines=true,
  frame=single,
  backgroundcolor=\color{gray!5},
  captionpos=b
}

\usepackage{xcolor}

\newif\ifhighlight
\highlightfalse 

\newcommand{\revise}[1]{%
  \ifhighlight
    \textcolor{red}{#1}%
  \else
    #1%
  \fi
}

\begin{document}

\title{Beyond Likes: How Normative Feedback Complements Engagement Signals on Social Media}


\author{Yuchen Wu}
\authornote{Both authors contributed equally to this research.}
\email{yuchen80@illinois.edu}
\affiliation{%
  \institution{University of Illinois Urbana-Champaign}
  \city{Urana}
  \state{Illinois}
  \country{USA}
}

\author{Mingduo Zhao}
\authornotemark[1]
\email{mingduo@berkeley.edu}
\affiliation{%
  \institution{UC Berkeley}
  \city{Berkeley}
  \state{California}
  \country{USA}
}

\author{John Canny}
\authornote{No Google authors used or analyzed Llama models for this work.}
\email{canny@berkeley.edu}
\affiliation{%
 \institution{UC Berkeley}
 \city{Berkeley}
 \state{California}
 \country{USA}
}
\affiliation{%
 \institution{Google LLC}
 \city{Mountain View}
 \state{California}
 \country{USA}
}

\renewcommand{\shortauthors}{Wu et al.}

\begin{abstract}

  \revise{Many online platforms incorporate engagement signals—such as likes—into their interface design to boost engagement. However, these signals can unintentionally elevate content that may not support normatively desirable behavior, especially when toxic content correlates strongly with popularity indicators. In this study, we propose structured prosocial feedback as a complementary signal, which highlights content quality based on normative criteria. We design and implement an LLM-based feedback system, which evaluates user comments based on principles from positive psychology, such as individual well-being. A pre-registered user study then examines how existing peer-based (popularity) and the new expert-based feedback interact to shape users' reposting behavior in a social media setting. Results show that peer feedback increases conformity to popularity cues, while expert feedback shifts choices toward normatively higher-quality content. This illustrates the added value of normative cues and underscores the potential benefits of incorporating such signals into platform feedback systems to foster healthier online environments.}
\end{abstract}




\maketitle

\section{Introduction}

Online platforms are designed to bring people together—to connect friends, facilitate conversations, and foster communities \cite{seraj2012we}. To promote such interaction at scale, most platforms rely heavily on popularity-based feedback mechanisms such as likes, shares, and retweets \cite{kim2018they,trunfio2021conceptualising}. Though simple in form, these signals exert significant influence—they offer users real-time feedback, elevate content visibility, and play a key role in how recommender systems prioritize content. As a result, they are central to the architecture of most social media platforms and serve as powerful drivers of user behavior.

These signals inherently influence user behavior and are far from neutral in their effects. Content that drives strong engagement often receives more positive feedback and greater visibility. However, such content may prioritize emotional appeal or entertainment value over substance—and in some cases, may even be toxic \cite{stieglitz2013emotions, berger2012makes}. This dynamic can undermine efforts to foster informative, inclusive, and reflective discourse. Empirical evidence suggests that popularity metrics systematically favor content that is emotionally intense, polarizing, or sensational—tending to elevate posts that incite reaction over those that promote constructive dialogue \cite{brady2020mad}. As a result, platforms face a fundamental tension: optimizing for engagement can sometimes amplify harmful behaviors or undermine the very sense of community they are meant to cultivate.

Traditional solutions to this problem have focused on content moderation \cite{gillespie2018custodians, chandrasekharan2018internet}. Harmful posts can be flagged, demoted, or removed; users may be warned or banned. While these methods are necessary for setting boundaries, they are fundamentally reactive, and they struggle to scale in real time. Moreover, they frame platform governance around suppression: around limiting what people shouldn't do, rather than nurturing what they could do \cite{einwiller2020online, srinivasan2019content}. They can feel punitive or authoritarian, triggering user resistance rather than reflection \cite{chandrasekharan2022quarantined}. 

In response, researchers and practitioners have begun exploring alternatives under the umbrella of proactive governance \cite{schluger2022proactive, habib2019act}: not just removing bad content, but also encouraging good behavior. This shift, from moderation to motivation, raises a key question: how might we design feedback mechanisms that not only discourage harmful behavior, but also encourage constructive, empathic, or civil contributions? While some platforms have begun to experiment with prompts and reminders \cite{kiskola2022online,yen2023storychat}, most continue to lack structured, value-aligned feedback tools that help users understand how their behavior maps onto broader social norms or community goals.

To fill this gap, we reimagine feedback not as a means of judgment or correction, but as a form of value-aware guidance. Drawing inspiration from movie sites like Rotten Tomatoes, which offer movie reviews from both audience members and critics (Experts), we explore how automatic (AI) feedback grounded in positive psychology principles can serve as a form of expert feedback. Can AI-generated feedback help humans reflect on and realign their behavior with community norms and platform values? Does this nudges human users toward more empathetic, constructive, or inclusive contributions?

Rather than replacing human judgment or enforcing strict norms, we propose a complementary feedback layer— interpretable, psychologically grounded, and designed to coexist with existing popularity-based metrics. Much like how reinforcement signals shape learning agents, our feedback aims to support human reflection in the moment of content creation. Through this lens, feedback becomes not a punitive or optimizing force, but a reflective mirror—a way for platforms to communicate their values softly, yet meaningfully, within everyday interactions.

To operationalize this idea, we designed and built a novel LLM-powered feedback system grounded in principles from positive psychology, which focuses on dimensions that support respectful, constructive, and beneficial conversation. Rather than offering detailed explanations or ranking users, the system assigns each post a single, structured numerical score from 0 to 10, where 0 indicates very poor alignment and 10 indicates very strong alignment with prosocial values, based on scoring rules we developed. The feedback is intentionally simple and unobtrusive, designed not to instruct but to prompt reflection, providing users with a new lens through which to interpret and evaluate their contributions.

To assess the behavioral impact of this approach, we conduct a controlled user experiment in which participants complete reposting tasks while receiving real-time prosocial feedback on candidate posts. Our study is designed to test how this new feedback signal interacts with existing traditional popularity cues, and whether it meaningfully shifts users’ preferences or attention. In particular, we examine whether users are willing to prioritize value-aligned content even when it receives fewer likes or retweets.

We designed four experimental conditions to evaluate different feedback mechanisms. The first was a control condition with no feedback. The second provided peer-based popularity feedback, mimicking likes and upvotes. The third introduced expert-based prosocial feedback, based on normative criteria. The final condition combined both popularity scores and expert scores, allowing us to examine their joint influence. We find that peer scores amplify conformity to popularity signals, often reinforcing status-quo engagement dynamics. In contrast, expert feedback—delivered as simple value-aligned scores, nudging users toward selecting more constructive and normatively desirable content, even when it lacks high popularity. When both signals are presented simultaneously, our findings highlight a key dynamic in the design of normative feedback: expert-based signals can meaningfully influence user behavior even in the presence of familiar peer-based cues. Specifically, when both popularity and normative scores were shown, users selected the comment rated higher by normative criteria 68.5\% of the time. In contrast, when only popularity feedback was available, this proportion dropped to just 50.9\% ($\Delta = +0.326$, $p = 0.000$). This suggests that normative signals can efficiently complement—rather than replace—existing popularity metrics, guiding users to more value-aligned content.

Together, these findings have both theoretical and practical implications. They suggest that value-aligned feedback can serve as a lightweight but meaningful form of proactive governance, helping users internalize platform norms not through coercion, but through subtle, reflective nudges. In sum, our contributions are threefold:


\begin{itemize}
    \item We propose the use of structured prosocial feedback as a motivational tool for behavior shaping on social platforms, offering a positive alternative to punitive moderation.
    
    \item We design and implement a feedback system scoring content based on positive psychology principles, and integrate it into a controlled user interface to study its impact on users' real-time decision-making.

    \item Through a pre-registered user experiment, we demonstrate that expert-based prosocial feedback can meaningfully shift user preferences toward more normatively aligned content—even when such content is less popular—highlighting its additive value beyond peer signals.

\end{itemize}

These contributions offer new insights into how online platforms might better align their behavioral incentives with their social values: not through suppression, but through encouragement. Designing feedback to convey not only engagement metrics but also the values we aim to promote creates a new design space. In this space, platforms serve not just as venues for connection, but also as subtle guides that shape how we interact, develop, and coexist within them.

\section{Related Work}

\subsection{Engagement-Oriented Design and Its Consequences}

Online platforms profoundly shape contemporary social interactions \cite{trunfio2021conceptualising, azzaakiyyah2023impact, olsson2020technologies}, not merely by hosting communication, but also by structuring how users see, interpret, and react to content within those environments. Central to this structure are engagement metrics—quantitative signals such as likes, shares, retweets, and view counts—which have become significant mechanisms for surfacing popular content and mediating collective attention in online platforms \cite{kim2018they,trunfio2021conceptualising, kim2021meaning}. These lightweight signals serve multiple functions: they help users gauge others’ opinions, allow content to propagate virally, and act as forms of social validation in online spaces. In doing so, they facilitate a scalable form of feedback that is deeply embedded in the platform interface and user experience.

However, engagement metrics are not neutral. Scholars have increasingly critiqued the implicit assumption that higher engagement equates to higher value or quality. In reality, optimizing for engagement can systematically advantage content that evokes strong emotional reactions, whether positive or negative. Empirical studies have demonstrated that emotionally arousing or morally charged content is more likely to be clicked, shared, and promoted by engagement-based algorithms \cite{milli2025engagement, vosoughi2018spread, brady2020mad}. 

These dynamics are further intensified by algorithmic curation, which draws heavily on engagement signals to determine what content is shown to whom. As a result, users are repeatedly exposed to highly engaged content, which can subtly distort collective perceptions of quality and desirability, privileging content that performs well under engagement-based metrics, regardless of its normative or informational value \cite{brady2023algorithm, milli2025engagement}.

Our work engages directly with this dilemma by asking: instead of eliminating engagement metrics, can we design complementary feedback systems that reshape user perception and behavior in more normatively desirable directions? We answer this question by exploring whether an alternative, prosocial feedback layer might offer users a different lens through which to evaluate and share content.

\subsection{Content Moderation and Platform Governance}

Content moderation is a cornerstone of platform governance, serving as a key mechanism through which platforms manage user behavior, enforce norms, and maintain community standards \cite{gillespie2018custodians, horta2023automated, roberts2016commercial, singhal2023sok}. Across a range of platforms—from mainstream services like Facebook and Reddit to niche online communities—moderation practices are primarily reactive: identifying and removing content that violates platform guidelines, issuing warnings or suspensions to users, or banning repeat offenders \cite{gillespie2018custodians, lampe2004slash, chandrasekharan2018internet}. These systems often use a combination of human moderators and automated tools, particularly in large-scale platforms such as Facebook, Reddit, and Twitter \cite{jhaver2019human, horta2023automated, edelson2023content}.

While such reactive and punitive mechanisms remain widespread, they face growing scrutiny. Scholars have highlighted several key limitations. First, users often perceive moderation decisions as opaque, inconsistent, or unfair, particularly when content is removed or accounts are penalized without adequate explanation \cite{jhaver2019did, gonccalves2023common}. This can lead to resistance, resentment, or even disengagement, especially when moderation is experienced as censorship rather than community care. Second, punitive models tend to focus on the suppression of harmful behavior rather than the cultivation of desirable alternatives. Simply removing problematic content does little to foster more inclusive, prosocial, or deliberative participation \cite{clune2024content, srinivasan2019content}.

In addition, punitive moderation may produce unintended consequences that undermine its intended goals. For example, scholars have documented how harsh enforcement can chill legitimate speech, drive users to migrate to less moderated platforms, or push harmful discourse into private or encrypted spaces where accountability is diminished \cite{gillespie2018custodians, russo2023spillover, horta2021platform}. These spillover effects highlight the limitations of content-focused interventions and suggest that effective governance must also consider the broader social dynamics of moderation and community engagement.

In response, researchers and practitioners have increasingly argued for a more nuanced, supportive, and proactive approach to platform governance—one that not only constrains harmful content but also nurtures community values and empowers users to make better choices. Our work is situated within this evolving paradigm, asking how we might augment traditional moderation with mechanisms that actively promote prosociality and reorient user attention toward community-beneficial content.

\subsection{Proactive and Value-Aligned Governance}

In recent years, scholars and practitioners have increasingly recognized the limitations of reactive moderation. Instead, there is growing interest in governance approaches that not only prevent harm but also actively promote positive behavior and social good \cite{gruning2024framework, schluger2022proactive, habib2019act}. These approaches shift attention from harm reduction toward cultivating healthier, more empathetic online environments.

A key design strategy in this space involves embedding social values directly into user-facing interfaces. For example, Yen et al. \cite{yen2023storychat} introduced StoryChat, a live-streaming chatroom tool that visualizes emotional narratives in real time. By making negative sentiment patterns visible, the system subtly nudged users toward more constructive and empathetic engagement, fostering a shared sense of community and mutual accountability. Similarly, Im et al. \cite{im2020synthesized} proposed Synthesized Social Signals (S3s), computational representations of a user’s behavioral history that are embedded in their profiles to promote transparency and help others calibrate their interactions. These tools exemplify how value-aligned interface design can scaffold prosocial participation and reduce the cognitive burden of social interpretation.

Another strand of research has focused on proactive moderation strategies that help assess and manage the emotional tenor of discussions in situ. Chang et al. \cite{chang2022thread} presented a tool that provides users with real-time feedback about tension levels in online threads as they compose replies. Rather than censoring content, the system encourages reflection and tone calibration, supporting users in avoiding unnecessary conflict. Schluger et al. \cite{schluger2022proactive} similarly analyzed proactive moderation practices on Wikipedia Talk Pages, finding that moderators often preemptively intervene in discussions to prevent escalation.

Collectively, these lines of work signal a broader paradigm shift: from governance as suppression to governance as cultivation. Our work builds upon this tradition by introducing a structured feedback layer grounded in value alignment. Drawing inspiration from work on prosocial nudges, we investigate how platforms might reinforce desirable behaviors not through punitive interventions, but through proactive intervention and feedback. In doing so, we join recent calls to rethink governance not as an exercise in control, but as an opportunity to cultivate the kinds of interactions that strengthen communities.

\subsection{Reimagining Feedback: From AI Training to Human Nudging}

Human behavior is profoundly influenced by feedback. Extensive research across psychology, behavioral economics, and human-computer interaction has demonstrated that structured feedback can significantly shape user behavior, attitudes, and decision-making processes \cite{gurjar2022effect}. For instance, studies have shown that positive social feedback for expressions of moral outrage increases the likelihood of future outrage expressions, consistent with principles of reinforcement learning \cite{brady2021social}. Similarly, researchers have demonstrated that reputation systems can significantly influence user behaviors in online communities \cite{adler2007content}. These findings suggest that feedback could serve as a formative signal, helping users calibrate their contributions in light of shared norms.

This formative potential of feedback is also recognized in community governance practice. In a recent study, Lambert et al. \cite{lambert2024positive} interviewed Reddit moderators and found that many deliberately provide positive feedback to encourage desired behavior and foster a more constructive discourse environment. Yet, despite its perceived efficacy, moderators noted a lack of explicit infrastructural support for such feedback, revealing an opportunity for design interventions that systematize and amplify positive reinforcement in online communities.

Our work builds on this premise but reimagines the design space. Inspired by existing systems like Rotten Tomatoes, which provides both peer (user ratings) and expert (Critic scores) reviews of movies, we use AI-based critics to encourage constructive posts instead of sanctioning less desirable ones. 

Our goal is to design feedback that is timely, interpretable, and ethically grounded. This positions feedback not as a corrective tool but as a nudge—a reflective mirror that invites users to consider alternative framings, more empathetic responses, or more inclusive perspectives, arguing for a reimagining of feedback: not as after-the-fact correction, but as real-time, value-aware social guidance.

\subsection{AI in Content Governance: Promise and Challenges}

Recent research has examined how AI systems can assist content governance by helping moderators identify harmful content, reduce manual workload, or surface decision-relevant signals \cite{jhaver2019human, horta2023automated, gillespie2020content, lai2022human}. These tools often aim to improve efficiency, consistency, or fairness in moderation decisions, offering scalable solutions for managing vast amounts of user-generated content. However, they also raise long-standing concerns regarding transparency, ethical issues, and the legitimacy of algorithmic authority in governing human expression \cite{gillespie2020content, nahmias2021oversight, molina2022ai, cortiz2020ethical}.

While prior systems often center on automating moderation decisions—deciding what content to remove, flag, or deprioritize—our work shifts the focus from enforcement to engagement. Rather than using AI to make determinations about appropriateness or compliance, we use it to generate simple, lightweight feedback grounded in shared prosocial values. These feedback signals do not constrain user behavior, nor do they act as justifications for moderation actions; instead, they are designed as gentle nudges—interpretable cues that help users reflect on the social and moral qualities of their contributions.

This approach is informed by a growing body of HCI research that views AI not only as an executor of policy but as a partner in shaping reflective, values-aligned digital spaces \cite{ehsan2020human}. By situating our work at the intersection of AI-assisted design and value-sensitive computing, we explore how feedback mechanisms can support a more participatory and human-centered model of content governance—one where behavioral shaping is not imposed, but co-constructed through subtle, value-driven signals.

\section{The Prosocial Feedback System}

To explore how structured feedback can gently guide user behavior toward more constructive and prosocial contributions, we developed an AI-powered feedback system designed to assess the "positive" psychological and social value of user-generated content, serving as an expert feedback from the platform named \emph{"Expert Score"}. Accordingly, we refer to traditional popularity-based feedback as \emph{"Peer Score"} in this paper, which is introduced as a baseline feedback signal in Section~\ref{sec:user_study}. 

Rather than relying on those popularity-based metrics like likes or shares, our system draws inspiration from positive psychology—a discipline focused on human flourishing, well-being, and character development. This section details how we translated the theoretical foundation of positive psychology into a working feedback mechanism, capable of evaluating social media posts through a nuanced, value-aligned lens.

\subsection{Theoretical Grounding: Three Perspectives on Positivity}

To define what “positive” means in the context of user contributions, we synthesized insights from three well-established perspectives in psychology and digital well-being.

\subsubsection{Individual Well-being Perspective}

We first grounded our approach in established frameworks from positive psychology that assess individual well-being. Specifically, we incorporated both the PERMA model and the Satisfaction with Life (SWL) scale to inform our AI-generated feedback mechanisms.

The PERMA model \cite{seligman2018perma} identifies five core elements contributing to human flourishing: Positive Emotion, Engagement, Relationships, Meaning, and Accomplishment. Each component offers a lens through which to assess the well-being implications of user content. 

In addition to PERMA, we utilized the Satisfaction with Life (SWL) scale \cite{diener1985satisfaction, pavot2008satisfaction}, a widely recognized measure of an individual's cognitive evaluation of their overall life satisfaction. Incorporating SWL allows for a broader assessment of well-being, capturing users' reflections on their life circumstances and overall contentment.

Computational studies have shown that linguistic cues in social media posts can predict subjective well-being across these dimensions \cite{schwartz2016predicting}. For example, a post reflecting gratitude or meaningful reflection might signal “Meaning” and “Positive Emotion,” while one that celebrates effort or persistence may reflect “Accomplishment”. By systematically evaluating these dimensions, our AI-generated feedback aims to provide users with insights into how their contributions resonate with aspects of personal well-being, encouraging reflection and the promotion of positive online interactions.

\subsubsection{Positive Social Media Use Perspective}

In addition to fostering individual well-being, social media platforms are deeply embedded in how people express themselves, maintain relationships, and participate in digital communities. To evaluate how user-generated content may reflect or encourage constructive and meaningful engagement online, we drew on three complementary strands of research: the Digital Flourishing Scale (DFS), empirical insights into the psychological benefits of social networking, and the Broaden-and-Build Theory of Positive Emotions.

The DFS, developed by Janicke-Bowles et al. \cite{janicke2023digital}, offers a comprehensive framework for evaluating positive perceptions of mediated social interactions. It encompasses five dimensions: authentic self-disclosure, civil participation, positive social comparison, connectedness, and self-control. These dimensions capture users' experiences of expressing their true selves, engaging respectfully in civic discourse, drawing inspiration from others, feeling socially connected, and exercising agency over their digital behaviors.

Complementing the DFS, a systematic review by Erfani and Abedin \cite{erfani2018impacts} identified key factors mediating the relationship between social network site usage and psychological well-being. These include perceived online social support, social capital, social self-esteem, authentic self-presentation, and social connectedness. Such factors underscore the potential of online interactions to bolster users' sense of belonging, self-worth, and community engagement.

Furthermore, the Broaden-and-Build Theory \cite{fredrickson2001role, fredrickson2004broaden} posits that positive emotions—such as joy, interest, contentment, and love—expand individuals' momentary thought-action repertoires, fostering creativity, exploration, and social bonding. Over time, these broadened mindsets build enduring personal resources, including social and psychological assets, which enhance individuals' capacity to cope with adversity and contribute to their overall well-being.

In operationalizing these frameworks, our AI system analyzes user content for features aligned with the aforementioned dimensions. For instance, narratives of personal growth or expressions of gratitude may signal authentic self-disclosure and positive emotion, while discussions of community involvement reflect civil participation and social capital. By providing feedback grounded in these social well-being constructs, our system aims to encourage users to engage in interactions that not only reflect their values but also foster a supportive and flourishing digital community.

\subsubsection{Character Strengths Perspective}

To further enrich our understanding of "positive" user contributions, we incorporated the Values in Action (VIA) classification of character strengths—a framework that identifies 24 universally valued traits such as kindness, curiosity, bravery, and humility \cite{ruch2010values}. These character strengths are not only morally aspirational but have also been empirically linked to various positive life outcomes, including academic achievement, healthy behaviors, mindfulness, life satisfaction, and reduced stress and depressive symptoms \cite{gillham2011character, gander2020character}. 

In our feedback system, we operationalized this perspective by analyzing user-generated content for indicators of these character strengths. For instance, narratives of overcoming challenges may reflect perseverance. This approach is informed by the work of Pang et al. \cite{pang2020language}, who demonstrated that language use on social media platforms like Twitter can reliably predict the presence of specific character strengths.

By integrating the VIA framework into our feedback mechanism, we aim to highlight and encourage the expression of these character strengths in online interactions. This not only promotes morally grounded behavior but also contributes to the development of a more supportive and constructive digital environment.

Combining these three frameworks, we created a multifaceted definition of positive content—one that goes beyond tone or civility to include psychological depth, social connectedness, and moral orientation.

\subsection{System Implementation: Chain-of-Thought Evaluation with GPT-4o}

To operationalize this multi-dimensional framework, we employed GPT-4o, a state-of-the-art large language model (LLM), as the engine behind our feedback system. We designed a chain-of-thought \cite{wei2022chain} prompting pipeline that guides the model through a structured, interpretable evaluation process.

First, we primed the model with descriptions and key examples from each of the three theoretical perspectives. After that, for each user-generated post, the model was asked to first understand user's post, and then reason explicitly about how the content aligned with each perspective. This involved generating brief justifications and assigning a 0–10 score for each dimension. 

After these individual assessments, the model was prompted to reflect on the three scores collectively and generate a composite score—again on a 0–10 scale. Importantly, we required the model to accompany this final score with a natural language explanation summarizing the rationale behind the evaluation.

This multi-step reasoning process offers several benefits:

\begin{enumerate}
    \item \textbf{Interpretability:} By making the model's reasoning transparent, we enhance trust and auditability in the feedback it generates.

    \item \textbf{Modularity:} Each scoring perspective is implemented as a separate module, enabling future versions of the system to easily incorporate new values or behavioral goals (e.g., sustainability, inclusivity).

    \item \textbf{Customizability:} The structured pipeline allows platforms to adjust weightings or dimensions to reflect their unique community norms and governance goals.
\end{enumerate}

To make the scoring process more intuitive, we present several examples in Table~\ref{table:score_example1}. A comment like “Excellent point. Funding for the GAO and more teeth around corruption would be a great way to improve things.” received a relatively high score of 7, as it demonstrates civic engagement, thoughtful problem-solving, and promotes prosocial discourse by offering encouragement to other viewpoints and valuing collective improvement. In contrast, a comment such as “The low-effort 'government bad, austerity is yes' spam is getting kinda old. Oh look, it came from the usual suspect.” received a low score of 3, due to its sarcastic tone, lack of constructive contribution, and minimal alignment with well-being principles such as positive emotion or kindness, and failure to enhance online social support or connectedness.

\begin{table}[h]
    \centering
    \caption{Scored Examples of User-Generated Content}
    \label{table:score_example1}
    \begin{tabular}{
        p{0.7\linewidth}
        p{0.13\linewidth}
        p{0.13\linewidth}
    }
         \toprule
         Question: Is there a huge waste in government spending? & \multicolumn{1}{c}{Peer Score} & \multicolumn{1}{c}{Expert Score} \\
         \midrule
         We need to be studying every one and figure out what works/doesn't work through real-life examples. Long term, a thoughtful public/private system is the goal, with real competition keeping costs in check and driving innovation, but with a well-designed public option that doesn't reward waste. & \multicolumn{1}{c}{5} & \multicolumn{1}{c}{7} \\ 
         \\
         The low-effort "government bad, austerity is yes" spam is getting kinda old.  Oh look, it came from the usual suspect. & \multicolumn{1}{c}{9} & \multicolumn{1}{c}{3} \\
         \\
         Did you not know there is only one government? Next, you are going to tell me there are more countries? Madness you are. & \multicolumn{1}{c}{8} & \multicolumn{1}{c}{2} \\ 
         \\
         Excellent point.  Funding for the GAO and more teeth around corruption would be a great way to improve things. & \multicolumn{1}{c}{4} & \multicolumn{1}{c}{7} \\
        \bottomrule
    \end{tabular}
\end{table}

\subsection{Design Philosophy and Practical Considerations}

Our goal in constructing this system was not to prescribe a universal definition of “good” content, but to demonstrate how value-driven feedback can be built in a transparent, extensible, and psychologically grounded way. While the backend system generates both a numerical score and an accompanying explanation to ensure transparency and interpretability during the feedback generation process, we deliberately chose to display only the score to users in this study. This decision was based on several practical and design considerations. By presenting a simple, familiar numerical signal—similar to likes or ratings—we aimed to reduce cognitive load and avoid overwhelming users with moral or psychological reasoning. At the same time, this abstraction allows the feedback to remain gentle and non-intrusive, encouraging reflection without prescribing behavior. The underlying explanations remain accessible for auditing or future expansion, preserving the system’s interpretability and extensibility for broader use cases.

\section{User Study: Exploring Responses to Different Feedback Signals}
\label{sec:user_study}

\begin{figure}[h]
  \centering
  \includegraphics[width=\linewidth]{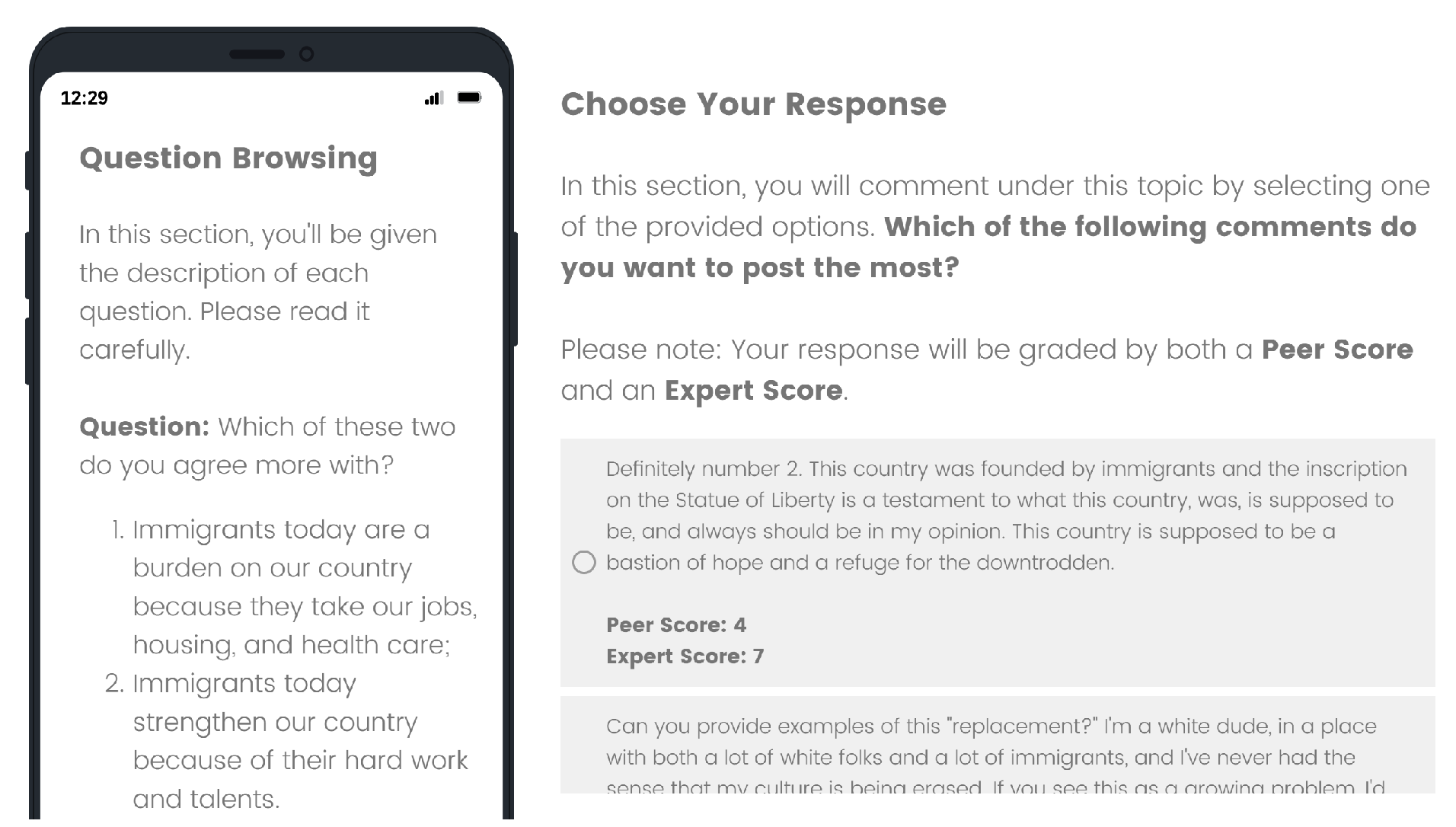}
  \caption{User interface of our comment-reposting simulation task. The left panel introduces the discussion topic question to participants. The right panel displays a set of pre-written comments, from which participants are asked to choose the one they would most like to repost. Each comment is accompanied by a Peer Score and an Expert Score in the presented condition, simulating feedback signals to investigate their influence on user choices.}
  \label{img:user_interface}
\end{figure}

To investigate the practical effects of our newly introduced feedback, we conducted a controlled user study simulating a comment-reposting experience on a social media platform. We created an interactive online experiment hosted on Qualtrics, which mirrors a social media environment. Participants were asked to choose which comment they would prefer to repost in response to a series of discussion questions (Fig~\ref{img:user_interface}). \revise{We opted for a reposting task rather than asking participants to write their own responses in order to ensure consistent effort across all possible choices and to enable controlled comparisons. This design allows us to isolate the effect of feedback on participants' judgments without introducing variance due to writing ability or motivation.} Crucially, the comments shown were drawn from the Reddit dataset containing real user responses and were pre-selected to reflect varying feedback signals, enabling us to observe user preferences under different feedback conditions. The study was preregistered at \href{https://aspredicted.org/msrh-xhqd.pdf}{https://aspredicted.org/msrh-xhqd.pdf}.

To detail our methodology, we break down the study into three components: the construction of feedback-driven comment sets, data sourcing and scoring, and the experimental procedure.

\subsection{Constructing Feedback-Driven Comment Sets}

In the user study, we simulated a feedback ecology with two kinds of evaluative signals: peer feedback, represented by normalized upvote counts drawn from Reddit data, and expert feedback, operationalized via our AI Feedback System that scores comments based on their alignment with principles from positive psychology we mentioned before.

We introduce both types of feedback for two key reasons:

\begin{enumerate}
    \item \textbf{To simulate the dual feedback ecology of real-world platforms:} Since social media platforms have long relied on popularity-based feedback, such as likes, upvotes, or shares, to gauge user engagement and influence content visibility, our newly introduced AI-generated, value-aligned feedback should be designed to coexist with these existing popularity-based signals. In this context, it is crucial to understand how users navigate the interplay between popularity-based \revise{(peer)} feedback and our prosocial \revise{(expert)} feedback. 

    \item \textbf{To examine behavioral responses in the presence of conflict:} Instead of examining expert feedback in isolation, we intentionally focused on conflict cases—situations where peer and expert feedback diverge. When both scores are high or both are low, individuals tend to either clearly prefer the high-scoring option or avoid the low-scoring one. By isolating conflict scenarios, we can better understand whether users are more inclined to follow the crowd or adopt platform-endorsed norms when the two diverge, especially when the AI feedback promotes a value that is not popular among peers.
\end{enumerate}

Therefore, the study was designed to answer the central question: When community feedback and AI-generated prosocial feedback conflict, which signal do users follow? Through this design, we aim to shed light on the social dynamics that may emerge as AI systems are increasingly used to steer behavior in online communities, not in isolation, but within the complex, sometimes contradictory landscape of social feedback.

\subsection{Data Preparation}

To ensure that the discussion questions in our study would resonate with users and elicit meaningful variation in response style, we selected eight socially relevant and value-laden topics (as shown in Table~\ref{table:topics}). These topics were drawn from reports by Pew Research Center \cite{pew2014polarization} and similar public opinion surveys, which identify domains of persistent disagreement and moral salience in online discourse, such as racial equity, climate change, education, and online privacy.

\begin{table}[t]
    \centering
    \caption{Topics Selected for the experiment}
    \label{table:topics}
    \begin{tabular}{
        p{0.34\linewidth}
        p{0.60\linewidth}
    }
         \toprule
         Topic & Question \\
         \midrule
         Government Waste & Is there a huge waste in government spending? \\
         \\
         Government Support for the Poor & Do you think increased government support for essentially a comfortable living would discourage the poor from working? \\
         \\
         Views on Immigration & Which of these two do you agree more with? (a) Immigrants today are a burden on our country because they take our jobs, housing, and health care; (b) Immigrants today strengthen our country because of their hard work and talents. \\
         \\
         US Military Spending & Should the US either actively use its military or significantly cut its defense spending? \\
         \\
         Government Regulation of Multinational Corporations & To what extent should governments regulate multinational corporations? \\
         \\
         Impact of Green Policies & Do green policies impact society positively or negatively? \\
         \\
         Black American Culture Issues & Will attributing certain social or cultural challenges to the Black American community influence real progress toward solutions? \\
         \\
         Regulation in Capitalism & Is regulation required for free markets/capitalism to work? \\
        \bottomrule
    \end{tabular}
\end{table}

For each topic, we scraped Reddit threads and curated a pool of user comments. Each comment was then assigned two scores:

\begin{itemize}
    \item \textbf{Peer Score:} Calculated from Reddit’s raw score (upvotes minus downvotes), normalized to a 0–10 scale within the thread.

    \item \textbf{Expert Score:} Computed using our AI Feedback System, which rates comments based on their alignment with positive psychology principles.
\end{itemize}

After this automated scoring process, we conducted a manual review of all candidate comments to ensure clarity, topical relevance, and diversity of perspectives. Our goal was not only to preserve linguistic fluency and coherence but also to construct a set of options that meaningfully represent tensions between popularity and psychological value.

For each discussion question, we selected four comments, all reflecting conflict scenarios where peer and expert scores diverged. Specifically, we included two types of contrast: One comment received many peer upvotes despite a low expert score (i.e., popular but lacking in substantive quality), while another was rated highly by the AI feedback system but received few peer upvotes (i.e., high-quality but socially unpopular).

To help illustrate these two types of scores, Table~\ref{table:score_example1} presents several examples. For instance, even though the comment “We need to be studying every one and figure out what works/doesn't work…” received a high Expert Score, its Peer Score was only 5, possibly because the language was dense and less engaging, making it harder to resonate with casual readers. On the other hand, the comment “Did you not know there is only one government?...” earned a much higher Peer Score of 8 despite scoring low on Expert criteria. Its playful rhetorical style and emotionally charged tone may have made it more appealing or entertaining to the online community, even if it lacked the prosocial depth valued by the expert framework.

Within each of these two categories, we included two distinct comments. This was a deliberate choice: by presenting participants with two comments of a similar evaluative profile but different content, we aimed to encourage deeper engagement. Rather than letting participants rely solely on a single surface-level cue (e.g., score or tone), we wanted them to read and reflect on the substantive qualities of the responses before making a choice. This subtle variation within each condition created opportunities for comparison, making the trade-off between popularity and normative value more salient and cognitively meaningful.

\subsection{Experimental Procedure}

In our study, each participant encountering four randomly selected questions out of the eight. Each question was paired with a different feedback condition and assigned in random order across participants. The four feedback conditions are defined by the feedback information shown:

\begin{enumerate}
    \item \textbf{No Feedback (Control):} No feedback signals are displayed.

    \item \textbf{Peer-Only Feedback:} Only peer feedback  is displayed.

    \item \textbf{Expert-Only Feedback:} Only expert feedback is displayed.

    \item \textbf{Dual Feedback:} Both peer and expert feedback are displayed.
\end{enumerate}



\begin{figure}[h]
  \centering
  \includegraphics[width=\linewidth]{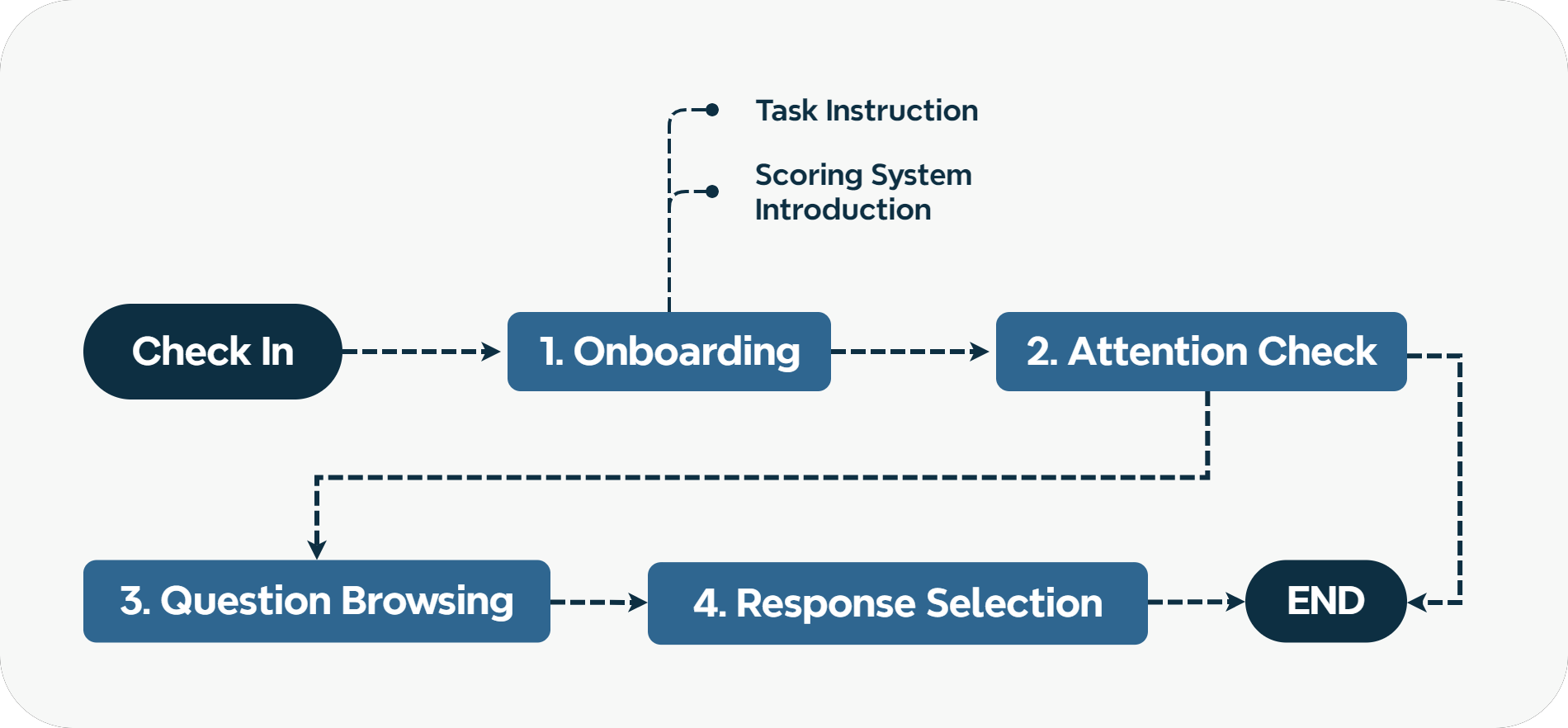}
  \caption{Main workflow diagram of the user study process.}
  \label{img:user_study_process}
\end{figure}

We recruited 546 U.S. participants via Prolific, of whom 496 passed the attention check and were included in our analysis. As shown in Fig~\ref{img:user_study_process}, participants first went through an onboarding phase, in which they were introduced to the overall structure of the study, including the decision task and the possible feedback mechanisms (no feedback, peer feedback, expert feedback, or both). \revise{In particular, participants were given detailed explanations of how the two types of feedback were generated: the Peer Score estimated the content's potential popularity among peers, while the Expert Score assessed the substantive quality of the content.} The purpose of this phase was to ensure participants had a clear understanding of the task and scoring systems before proceeding.

Following onboarding, participants completed an attention check designed to verify engagement and task comprehension. This included two multiple-choice tasks requiring them to identify key instructions they had just seen. Participants who failed the attention check were excluded from the remainder of the study.

Participants who passed the check proceeded to the main experiment. Each participant was presented with four topic questions one by one. Within each topic, participants viewed the four curated comments (presented in randomized order) and answered the following question:

\textit{“You will comment under this topic by selecting one of the provided options. Which of the following comments do you want to post the most?”}

Participants were instructed to select the comment they most preferred to post, with no right or wrong answers. \revise{In this setup, participants did not generate new responses; instead, they effectively ‘reposted’ one of the existing comments as their response to the topic question. We use the Peer Score and Expert Score of this response in our analysis that follows.} After completing the four main decision tasks, participants answered a short questionnaire regarding their demographic information.

\section{Results}

We present the results of our user study in two parts. First, we examine the effect of different feedback types on participant behavior by analyzing mean differences across conditions. Specifically, our primary dependent variables are the Peer Score and Expert Score of the comments participants selected in each condition, which capture the popularity and normative quality of their choices, respectively. Second, we validate the robustness of these findings.

\subsection{Mean Differences Across Feedback Conditions}

\begin{figure}[t]
  \centering
  \includegraphics[width=\linewidth]{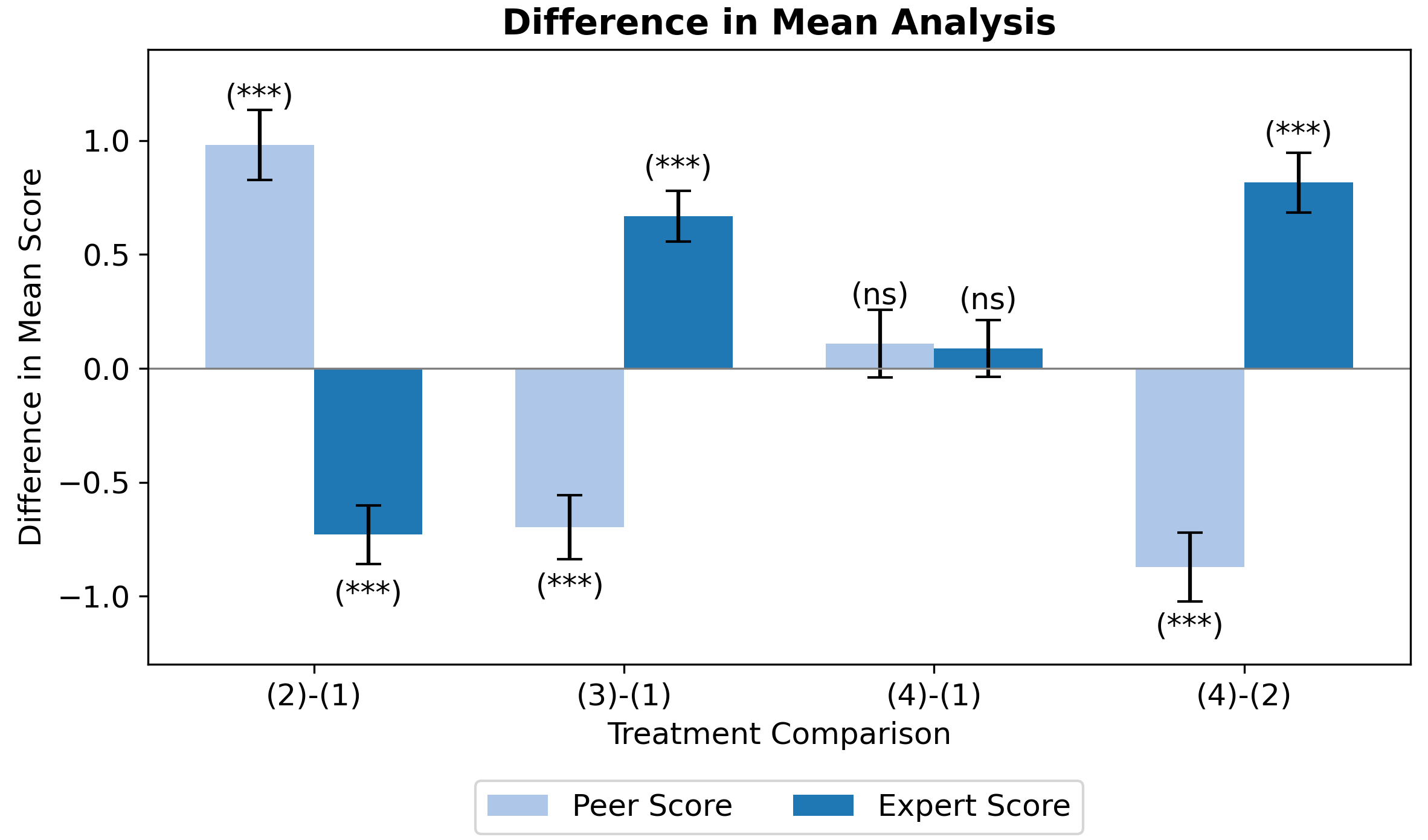}
  \caption{Results for the difference in mean analysis. Bars represent the differences in mean peer and expert scores between condition pairs. Conditions (1) through (4) correspond to \textit{No Feedback}, \textit{Peer-Only Feedback}, \textit{Expert-Only Feedback}, and \textit{Dual Feedback}, respectively. Error bars indicate the standard error of the mean difference. The symbols in parentheses show the corresponding p-value. *** for p < 0.01, ** for p < 0.05, * for p < 0.1. Non-significant results are denoted as "ns".}
  \label{img:diff_in_mean}
\end{figure}

We began by examining the average Peer Score and Expert Score of selected comments across each condition (Table~\ref{table:diff_in_mean}, Fig~\ref{img:diff_in_mean}). In the control condition (no feedback), participants selected comments with moderate peer ($M=5.008$) and expert scores ($M=5.637$). As expected, Peer-only feedback (Condition 2) led to higher Peer Scores ($M=5.990)$, but at the cost of lower Expert Scores ($M=4.907$). On the contrary, Expert-only feedback (Condition 3) increased Expert Scores ($M=6.305$), but decreased Peer Scores ($M=4.311$). Interestingly, in the Dual condition, both Peer and Expert Scores of selected comments increased ($M=5.118$ and $5.724$, respectively).

To quantify these effects, we go beyond simple pairwise comparisons by computing differences in means. Relative to the control group, Peer-only feedback significantly increased Peer Scores ($\Delta = +0.981$, $p = 0.000$) but decreased Expert Scores ($\Delta = -0.730$, $p = 0.000$), indicating that participants gravitated toward popular comments when exposed to popularity cues. In contrast, Expert-only feedback significantly increased Expert Scores ($\Delta = +0.668$, $p = 0.000$) while lowering Peer Scores ($\Delta = -0.697$, $p = 0.000$), showing a stronger alignment with normative goals. The Dual feedback condition, however, did not produce significant changes in either Peer Score ($p = 0.460$) or Expert Score ($p = 0.484$) compared to the control. While the difference from the control condition was not statistically significant, this suggests a partial balancing effect: when both scores are visible, users may integrate both types of feedback, trying to make choices that are reasonably popular and also normatively constructive.

Crucially, when we compare the Dual feedback condition directly with the Peer-only condition—the platform baseline we seek to improve—we observe a significant drop in Peer Score ($\Delta = -0.872$, $p = 0.000$) and a significant rise in Expert Score ($\Delta = +0.816$, $p = 0.000$). This suggests that supplementing peer-based feedback with expert scores can meaningfully shift user behavior toward more normatively desirable responses. These findings support the idea that expert-based signals can act as a corrective force, tempering the influence of peer-driven engagement metrics in Dual feedback settings and nudging users toward more normatively aligned participation.

\begin{table}[t]
    \centering
    \caption{Results of the Difference in Means Analysis}
    \label{table:diff_in_mean}
    \begin{tabular}{lcccc}
         \toprule
         & (1) & (2) & (3) & (4) \\
        Condition & No Feedback & Peer-Only & Expert-Only & Dual \\
        & (Control) & Feedback & Feedback & Feedback \\
        \midrule
        Peer Score & 5.008 & 5.990 & 4.311 & 5.118 \\
        Expert Score & 5.637 & 4.907 & 6.305 & 5.724 \\
        \bottomrule
    \end{tabular}
    \\
    \vspace{8 pt}
    \begin{tabular}{lcccc}
        \toprule
        & (5) & (6) & (7) & (8) \\
        Condition Pair & (2)-(1) &(3)-(1) & (4)-(1) &(4)-(2) \\
        \midrule
        Peer Score & 0.981 & -0.697 & 0.109 & -0.872 \\
        & (0.000) & (0.000) & (0.460) & (0.000) \\
        Expert Score & -0.730 & 0.668 & 0.087 & 0.816 \\
        & (0.000) & (0.000) & (0.484) & (0.000) \\
        \bottomrule
    \end{tabular}
    \\
    \vspace{25 pt}
     \textbf{Notes:}  The number in parentheses is the corresponding p-value obtained from the t-test.
\end{table}

\subsection{Robustness of Behavioral Effects Across Conditions}

To ensure the robustness of our results and rule out the possibility that observed behavioral differences arise from speciﬁcation choices, we conducted additional analyses using two complementary approaches: a proportion-based preference metric and non-parametric permutation tests.

\subsubsection{Preference for Expert-Endorsed Comments}

To complement the mean score analysis, we examined the proportion of users in each condition who chose the comment with the higher expert-assigned score over the one with the higher peer-assigned score (Table~\ref{table:diff_in_proportion}). This allows us to assess what proportion of users prioritize normative quality over popularity when the two are in conﬂict. The proportion-based measure provides a clear summary of users' relative preference for expert-endorsed versus peer-endorsed comments.

As expected, the proportion was highest in the Expert-only condition (83.5\%) and lowest in the Peer-only condition (50.9\%), with the control and Dual feedback conditions in between (67.2\% and 68.5\%, respectively). Compared to the control, Expert-only feedback significantly increased this preference ($\Delta = +0.163$, $p = 0.000$), while Peer-only feedback significantly reduced it ($\Delta = -0.163$, $p = 0.000$), echoing the trends observed in average selection scores. Importantly, participants in the Dual feedback condition were significantly more likely to prefer expert-endorsed comments than those in the Peer-only condition ($\Delta = +0.326$, $p = 0.000$), reinforcing our conclusion that expert signals help correct for peer influence and promote alignment with platform norms.

\begin{table}[b]
    \centering
    \caption{Proportion of Users Preferring Expert Feedback Over Peer Feedback}
    \label{table:diff_in_proportion}
    \begin{tabular}{lcccc}
         \toprule
         & (1) & (2) & (3) & (4) \\
        Condition & No Feedback & Peer-Only & Expert-Only & Dual \\
        & (Control) & Feedback & Feedback & Feedback \\
        \midrule
        Proportion & 0.672 & 0.509 & 0.835 & 0.685 \\
        \bottomrule
    \end{tabular}
    \\
    \vspace{8 pt}
    \begin{tabular}{lcccc}
        \toprule
        & (5) & (6) & (7) & (8) \\
        Condition Pair & (2)-(1) &(3)-(1) & (4)-(1) &(4)-(2) \\
        \midrule
        Proportion & -0.163 & 0.163 & 0.012 & 0.326 \\
        & (0.000) & (0.000) & (0.731) & (0.000) \\
        \bottomrule
    \end{tabular}
    \\
    \vspace{8 pt}
     \textbf{Notes:}  The number in parentheses is the corresponding p-value obtained from the Chi-Square test.
\end{table}

\subsubsection{Permutation Test Validation}

To further assess statistical robustness, we conducted permutation tests for all pairwise comparisons between experimental conditions (Table~\ref{table:permutation_test_peer} and Table~\ref{table:permutation_test_expert}). These tests confirmed the key results from the mean-difference analyses. Most condition pairs yielded statistically significant differences ($p = 0.000$), except for comparisons involving the Dual and Control groups, which consistently produced non-significant results. This further validates the observation that Dual feedback does not reliably alter behavior compared to baseline.

\begin{table}[h]
    \centering
    \caption{Permutation Test Results (p-values) for Peer Scores}
    \label{table:permutation_test_peer}
    \begin{tabular}{lcccc}
         \toprule
         & No Feedback & Peer-Only & Expert-Only & Dual \\
         & (Control) & Feedback & Feedback & Feedback \\
         \midrule
         No Feedback (Control) & - & 0.000 & 0.000 & 0.463 \\
         Peer-Only Feedback & 0.000 & - & 0.000 & 0.000 \\
         Expert-Only Feedback & 0.000 & 0.000 & - & 0.000 \\
         Dual Feedback & 0.463 & 0.000 & 0.000 & - \\
        \bottomrule
    \end{tabular}
\end{table}

\begin{table}[h]
    \centering
    \caption{Permutation Test Results (p-values) for Expert Scores}
    \label{table:permutation_test_expert}
    \begin{tabular}{lcccc}
         \toprule
         & No Feedback & Peer-Only & Expert-Only & Dual \\
         & (Control) & Feedback & Feedback & Feedback \\
         \midrule
         No Feedback (Control) & - & 0.000 & 0.000 & 0.495 \\
         Peer-Only Feedback & 0.000 & - & 0.000 & 0.000 \\
         Expert-Only Feedback & 0.000 & 0.000 & - & 0.000 \\
         Dual Feedback & 0.495 & 0.000 & 0.000 & - \\
        \bottomrule
    \end{tabular}
\end{table}

\section{Discussion}

\subsection{Peer Feedback Encourages Popularity-Conforming Choices}

Our results show that displaying Peer Scores alone significantly increased the likelihood that users selected comments with higher peer endorsement. This finding aligns with a substantial body of prior work in social computing and behavioral economics demonstrating that popularity-based signals—such as likes, upvotes, and user ratings—can powerfully steer attention, preferences, and behavior in digital environments \cite{muchnik2013social, merchant2019signals}. Even in our minimalist implementation, where peer feedback was shown via a simple numerical score, participants reliably gravitated toward content with higher peer approval. This suggests that social proof remains a dominant heuristic in user decision-making, even when no additional context or explanation is provided.

From a psychological perspective, this behavior reflects the broader cognitive tendency to conform to perceived majority opinion, a dynamic that is especially amplified in online spaces where ambiguity is high and normative cues are scarce \cite{smith2007uncertainty, asch1955opinions, noelle1974spiral}. The presence of peer endorsement reduces the perceived risk of making a "bad" choice and can serve as a shortcut for quality assessment when users lack the time or motivation to deeply evaluate content themselves. In our study, the presence of peer scores likely lowered users’ threshold for information scrutiny, causing them to lean more heavily on consensus cues.

This mechanism is particularly important for online platforms that aim to foster community participation. Peer-based feedback can serve as a powerful tool to highlight valued contributions, motivate content creators, and streamline information discovery by surfacing widely endorsed content \cite{shankar2024nonverbal, tu2019feedback}. In this respect, our findings validate design practices commonly implemented across social computing systems that leverage engagement metrics as proxies for quality or relevance.

However, the strength of this mechanism also surfaces important concerns. Prior research has warned that popularity signals can lead to rich-get-richer effects, where early upvotes snowball into disproportionate visibility, reinforcing existing biases rather than surfacing the most informative or diverse content \cite{ciampaglia2018algorithmic, zhu2021popularity, ciampaglia2018algorithmic}. Our findings echo these concerns by showing that peer feedback alone may encourage conformity to prevailing opinion, possibly at the expense of 
positive psychological and social value. Over time, such dynamics can entrench echo chambers and narrow the range of visible perspectives on a platform, highlighting the need for careful calibration in feedback system design.

\subsection{Expert Feedback Nudges Normative Selection Behavior}

Our findings show that introducing Expert Scores significantly increased the likelihood that users selected comments with higher normative quality. This effect demonstrates that users are not solely driven by popularity cues but are also responsive to signals that encode content quality or epistemic value, which are difficult to surface through peer endorsement alone. Unlike popularity signals, which may amplify social consensus regardless of content substance, expert signals offer an effective channel for shaping user behavior toward higher-quality information selection. As such, expert feedback mechanisms could be especially valuable in contexts where misinformation, incivility, or low-effort contributions are common, such as online debates, comment sections, or knowledge-sharing forums.

Interestingly, this behavioral shift occurred without any elaboration on the identity or credentials of the “experts”. Participants were simply told that the score reflected an expert evaluation, without further justification. This suggests that even minimal expert framing can invoke a sense of authority or credibility sufficient to influence user behavior. In line with prior literature on heuristic processing and source credibility \cite{chaiken1994heuristic, metzger2013credibility, metzger2010social}, users may treat expert scores as cognitively efficient signals for content quality, particularly in fast-paced or information-rich settings where careful deliberation is costly.

Therefore, our findings raise important questions about the design and governance of expert-based interventions. First, while users responded to the presence of expert feedback, it remains unclear how this influence would evolve over time, especially if users became skeptical of the expert source or perceived it as ideologically biased. Second, the persuasive power of an “expert” label—even when not supported by transparent explanations—poses potential risks if misused. As algorithmic systems increasingly simulate or replace traditional expert judgments (e.g., via AI-generated scores), ensuring fairness, accountability, interpretability, and legitimacy becomes critical.


\subsection{Layering Expert Feedback on Top of Peer Feedback Adds Value}

In most real-world environments, platforms do not operate in a vacuum—they already embed rich ecosystems of peer-based metric. Thus, understanding how normative signals like expert feedback interact with these entrenched popularity cues is critical. Our comparison of the Dual condition (where both Peer and Expert Scores were shown) against the Peer-only condition simulates this layered environment. 

The significant increase in Expert Score under the Dual condition indicates that expert feedback can work within existing design paradigms—it was noticed, processed, and incorporated into user decision-making. There is no need for a radical overhaul of peer-based infrastructures to begin reorienting user behavior toward more normatively desirable outcomes. Instead, platforms can take a layered, incremental approach: retain peer signals to preserve familiarity and user engagement, while gradually introducing expert cues to subtly shift attention and judgment criteria. Such layering not only eases adoption from the user perspective but also reduces product risk for platform designers who may be hesitant to disrupt high-engagement social features.

Moreover, this interaction enables a potential complementarity between popularity and quality signals. Peer-based endorsements can serve as rough indicators of community interest or salience, while expert feedback injects a quality filter aligned with epistemic or normative goals. By combining the two, platforms may create more balanced recommendation environments—ones that reward both crowd resonance and substantive merit. This hybrid model could be especially beneficial in domains where both popularity and accuracy matter, such as health forums, news comments, or educational Q\&A sites.

Overall, these findings encourage a pragmatic design philosophy: rather than forcing a binary choice between peer-driven and expert-driven architectures, platforms can combine the strengths of both. Even subtle, well-integrated additions of expert feedback can meaningfully elevate the quality of user selections—without alienating users or disrupting engagement dynamics.

\subsection{When More Is Not Always Enough: The Challenges of Mixed Feedback}

Indeed, we found that adding expert feedback on top of peer feedback led to more normatively aligned selections (as evidenced by higher Expert Scores compared to Peer-only feedback condition). However, when comparing the Dual condition to the Control group (i.e., no feedback at all), this advantage diminished—a somewhat counterintuitive result that deserves closer examination.

One plausible explanation lies in signal conflict and cognitive ambiguity. In our trials, the peer and expert scores pointed to different comments—some highly endorsed by peers received low expert ratings, and vice versa. Without clear framing or prioritization cues, some users may have found it difficult to resolve this tension, resulting in indecision, confusion, or even complete disengagement from the feedback. This could explain why, on average, users in the Dual condition did not make significantly more aligned selections than those who saw no feedback at all.  Additionally, the absence of explanations about who the experts were or how the scores were derived likely limited users’ ability to meaningfully integrate the signals.

Design-wise, these findings suggest that simply adding multiple feedback channels is insufficient. Appropriate explanations, interface framing, and visual hierarchy might be beneficial for helping users resolve conflicts and extract value from feedback. In particular, future designs could explore ways to embed light-touch explanations—for example, short phrases like “Experts rated this highly for constructive suggestions and active civic participation”—which can ground each score in a meaningful rationale, helping users make more informed choices when signals diverge. Progressive disclosure—initially showing only one type of score (e.g., expert feedback), with the option to reveal peer ratings on demand—may be helpful, too. It can reduce information overload and give users greater control over how much evaluative input they consider. Visual prioritization can also play a key role: emphasizing expert scores more prominently (e.g., via color, size, or layout) may help guide user attention toward normatively desirable content. Finally, explicit conflict alerts—such as “Expert and peer feedback differ—click to see why”—could help users recognize when signals are in tension and invite deeper engagement, rather than passive confusion.

In sum, with supportive framing and explanation, mixed feedback may realize its full potential in guiding user behavior.

\subsection{General Implications}

This work contributes to a growing conversation about how platforms can move beyond engagement-centric designs toward value-aligned interfaces, ones that encourage thoughtful participation, elevate high-quality contributions, and balance social and normative cues. Our results suggest that platforms do not necessarily need to abandon familiar peer-driven mechanisms to introduce expert guidance. Instead, they can incrementally enrich the feedback environment by integrating expert signals in ways that complement, rather than disrupt, existing social signals. By surfacing both the opportunities and tensions of mixed feedback design, our study provides a foundation for building platforms that support more informed, reflective, and socially constructive online participation.

\subsection{Limitations and Future Work}

\revise{
Our work has several limitations that point toward promising avenues for future work. First, the study was conducted in a controlled lab setting rather than in real-world settings. While this design enabled systematic comparisons across conditions, it may not fully reflect the dynamics of posting behaviors on live social media platforms. Deployments in the field, such as A/B tests on actual platforms, would help determine whether the observed effects generalize at scale and under real-world constraints. Second, our study focused on short-term behavioral outcomes, leaving open whether effects persist over time or shape longer-term attitudes and trust. Longitudinal studies are needed to assess the durability of these influences. Finally, our design presented expert feedback as a simple numerical score without explanations. Future designs could explore richer presentation formats, such as narrative rationales, confidence indicators, or value-oriented prompts, to help users interpret and integrate signals more effectively.
}


\section{Conclusion}

This study investigated how different forms of evaluative feedback—peer-based, expert-based, and a combination of both—shape user reposting behavior in online content environments. Our findings reveal that even lightweight feedback signals can meaningfully influence user choices: peer feedback promotes conformity to popularity, expert feedback nudges toward normatively strong content, and layering expert cues atop peer signals can augment their impact.

These insights have direct implications for the design of digital platforms seeking to promote higher-quality engagement. Rather than relying solely on engagement metrics, platforms can embed expert-aligned feedback in ways that guide users toward more thoughtful or prosocial contributions. Our work also suggests that thoughtful layering, contextual framing, and explanatory design elements could play an important role in unlocking the full potential of mixed feedback systems.

More broadly, our findings point to new possibilities for proactive content moderation and platform governance. Instead of relying exclusively on reactive removals or manual reviews, platforms could shift toward nudging mechanisms that guide user behavior upstream, prior to harmful or low-quality content gaining traction. Expert-informed feedback, presented with care, offers one such scalable and interpretable intervention. Integrating such approaches into content recommendation, comment ranking, or participatory review systems may help platforms balance openness with normative goals like truthfulness, civility, and deliberative quality.

By clarifying how different types of feedback shape user decisions, this research lays a foundation for developing more value-aligned, cognitively supportive social systems. \revise{Future work could build on these findings through longitudinal and real-world evaluations, as well as by exploring alternative definitions and designs of normative feedback.}


\bibliographystyle{ACM-Reference-Format}
\bibliography{sample-base}

\appendix
\UseRawInputEncoding

\section{LLM Prompts}

\begin{lstlisting}[style=promptblock]

You are an expert in positive psychology tasked with evaluating the quality of social media posts/comments from a well-being perspective. Analyze the given content using the following framework:

### 1. Individual Well-being Prediction:

- Assess how the content reflects:
    1. Satisfaction with Life (SWL): a well-established representation of well-being, representing aperson’s cognitive evaluation of their own life.
    2. PERMA model elements (Positive emotion, Engagement, Relationships, Meaning, and Accomplishment). 
        - Positive emotion includes positively valenced emotions such as joy, contentment, and excite-ment. 
        - Engagement is a multi-dimensional construct that includes behavioral, cognitive, and affec-tive components. It can refer to involvement and participation in groups or activities, enthusiasm andinterest in activities, commitment and dedication to work, and focused attention to tasks at hand. For our purposes here, we deﬁne it in terms of passion and involvement in life, as opposed to apathyand boredom. 
        - Relationships (or positive relationships) includes trusting others, perceiving others asbeing there if needed, receiving social support, and giving to others. Considerable evidence identiﬁes the importance of positive relationships for supporting health, longevity, and other importantlife outcomes.
        - Meaning in life captures having a sense of purpose, signiﬁcance, and understandingin life. It can also include transcending the self, feeling a sense of connection to a higher poweror purpose, and provides goals or a course of direction to follow.
        - Accomplishment is often deﬁnedin terms of awards, honors, and other objective markers of achievement. For our purposes here,we focus on the subjective side, in terms of a personal sense of accomplishment. It includes a senseof mastery, perceived competence, and goal attainment.

### 2. Social Media Benefits:

- Evaluate the content based on the Digital Flourishing Scale (DFS) dimensions:
    a. Authentic self-disclosure
    b. Civil participation
    c. Positive social comparison
    d. Connectedness
    e. Self-control
- Analyze how the post/comment might contribute to or reflect:
    f. Perceived online social support
    g. Social capital
    h. Social self-esteem
    i. Social connectedness
- Consider how the content aligns with the Broaden-and-Build Theory of Positive Emotions:
    a. broaden an individual's momentary thought–action repertoire: joy sparks the urge to play, interest sparks the urge to explore, contentment sparks the urge to savour and integrate, and love sparks a recurring cycle of each of these urges within safe, close relationships.
    b. build that individual's personal resources: ranging from physical and intellectual resources, to social and psychological resources. Importantly, these resources function as reserves that can be drawn on later to improve the odds of successful coping and survival.

### 3. Character Strengths:

- Refer to the list of 24 VIA Character Strengths attached, and identify any of the 24 VIA Character Strengths that are evident in the content.
- Assess how these strengths contribute to academic achievement, healthy behaviors, mindfulness, life satisfaction, multi-dimensional well-being, and orientation to happiness, improving well-being and reducing depressive symptoms and stress

### 4. Additional Considerations:

- Reflect on any other aspects of positive psychology or well-being that are relevant to the content but not covered in the above categories.

### 5. Scoring:
- Based on your analysis of these categories, please provide scores for each sub-category from 0 to 10.
- And then integrate these scores to provide an overall evaluation of the comment's potential popularity, by assigning a final score from 0 to 10.
- 0 represents the lowest popularity, while 10 represents the highest popularity.
- Provide an overall explanation of your final score, highlighting the key factors that influenced your evaluation. Your explanation should be formal, professional, and analytical. Ensure your language is precise, objective, and academically rigorous. The explanation should provide a detailed, structured analysis of the comment's qualities, while maintain the concise and focused nature of a professional evaluation.


Here is the information for your evaluation:

Comment: <COMMENT>

Please respond in the following JSON format:
```json
{
    "step1_individual_well_being": {
        "score": <SCORE>,
        "explanation": "Provide a brief explanation of the score for this category."
    },
    "step2_social_media_benefits": {
        "score": <SCORE>,
        "explanation": "Provide a brief explanation of the score for this category."
    },
    "step3_character_strengths": {
        "score": <SCORE>,
        "explanation": "Provide a brief explanation of the score for this category."
    },
    "step4_additional_aspects": <COMMENTARY>,
    "step5_overall_thoughts": <OVERALL_THOUGHTS>,
    "step6_final_score": <FINAL_SCORE>
}

[MATERIAL] Values in Action Classification of the 24 Strengths of Character
Virtue I. Wisdom and knowledge: cognitive strengths that entail the acquisition and use of knowledge.
    (1) Creativity (originality, ingenuity): thinking of novel and productive ways to conceptualize and do things; includes artistic achievement but is not limited to it
    (2) Curiosity (interest, novelty seeking, openness to experience): taking an interest in all of ongoing experience for its own sake; finding subjects and topics fascinating; exploring and discovering
    (3) Judgement (open-mindedness, critical thinking): thinking things through and examining them from all sides; not jumping to conclusions; being able to change one’s mind in light of evidence; weighing all evidence fairly
    (4) Love of learning: mastering new skills, topics, and bodies of knowledge, whether on one’s own or formally; obviously related to the strength of curiosity, but goes beyond it to describe the tendency to add systematically to what one knows
    (5) Perspective (wisdom): being able to provide wise counsel to others; having ways of looking at the world that make sense to oneself and to other people
Virtue II. Courage: emotional strengths that involve the exercise of will to accomplish goals in the face of opposition, external or internal.
    (6) Bravery (valor): not shrinking from threat, challenge, difficulty, or pain; speaking up for what is right even if there is opposition; acting on convictions even if they are unpopular; includes physical bravery but is not limited to it
    (7) Perseverance (persistence, industriousness): finishing what one starts; persisting in a course of action in spite of obstacles; “getting it out the door”, taking pleasure in completing tasks
    (8) Honesty (integrity, authenticity): speaking the truth and presenting oneself in a genuine way; being without pretense; taking responsibility for one’s feelings and actions
    (9) Zest (vitality, vigor, enthusiasm, energy): approaching life with excitement and energy; not doing things halfway or halfheartedly; living life as an adventure; feeling alive and active
Virtue III. Humanity: interpersonal strengths that involve ‘‘tending and befriending’’ others.
    (10) Love: valuing close relations with others, in particular those in which sharing and caring are reciprocated; being close to people
    (11) Kindness (generosity, nurturance, care, compassion, altruistic love, “nice-ness”): doing favors and good deeds for others; helping them; taking care of them
    (12) Social intelligence (emotional intelligence, personal intelligence): being aware of the motives and feelings of other people and oneself; knowing what to do to fit into different social situations; knowing what makes other people tick
Virtue IV. Justice: civic strengths that underlie healthy community life.
    (13) Teamwork (citizenship, social responsibility, loyalty): working well as member of a group or team; being loyal to the group; doing one’s share
    (14) Fairness: treating all people the same according to notions of fairness and justice; not letting personal feelings bias decisions about others; giving everyone a fair chance
    (15) Leadership: encouraging a group of which one is a member to get things done, and at the same time maintain good relations within the group
Virtue V. Temperance: strengths that protect against excess. 
    (16) Forgiveness (mercy): forgiving those who have done wrong; accepting the shortcomings of others; giving people a second chance; not being vengeful
    (17) Modesty (humility): letting one’s accomplishments speak for themselves; not seeking the spotlight; not regarding oneself as more special than one is
    (18) Prudence: being careful about one’s choices; not saying or doing things that might later be regretted
    (19) Self-regulation (self-control): regulating what one feels and does; being disciplined; controlling one’s appetites and emotions
Virtue VI. Transcendence: strengths that forge connections to the larger universe and provide meaning.
    (20) Appreciation of beauty and excellence (awe, wonder, elevation): noticing and appreciating beauty, excellence, and/or skilled performance in all various domains of life, from nature, to art, to mathematics, to science, to everyday experience.
    (21) Gratitude: being aware of and thankful for the good things that happen; taking time to express thanks
    (22) Hope (optimism, future-mindedness, future orientation): expecting the best and working to achieve it; believing that a good future is something that can be brought about
    (23) Humor (playfulness): liking to laugh and tease; bringing smiles to other people; seeing the light side; making (not necessarily telling) jokes
    (24) Spirituality (religiousness, faith, purpose): having coherent beliefs about the higher purpose and meaning of the universe; knowing where one fits within the larger scheme; having beliefs about the meaning of life that shape conduct and provide comfort

\end{lstlisting}

\end{document}